# LEARNING ABOUT THE REDUCTION OF FOOD WASTE USING BLOCKCHAIN TECHNOLOGY

**Monica-Paula Marin[1], Iuliana Marin[2], Livia Vidu[1]**

[1]*University of Agronomic Sciences and Veterinary Medicine of Bucharest, Romania*
[2]*University Politehnica of Bucharest, Romania*

## Abstract

Farmers need to be efficient and dedicate a lot of time in order to sustain the quality of their animals which are in their care. The most convenient and good quality - price ratio should be chosen for the feed of animals. Blockchain is used in a virtual space to store and share information over a network of users. This is done using the open source Hyperledger Fabric platform. The transactions can be viewed by all the other users in real time. These transactions are stored as JSONs inside CouchDB NoSQL database which supports queries on a large volume of data. When using this technology, the farmer can know with whom the supplier for animal feed collaborated with. The history of the transactions are not saved in just one place. In this way, it is more difficult to hack and provide implausible information.

An e-learning platform was created where the farm's user can post information, respectively new blocks about the animal's birth, vaccinations, medicines, including the location of the livestock. The same e-learning platform is accessible from the mobile phone. By using the blockchain technology, anyone, including the client from the shop can know a lot about the origin of the products. Fake origins of food are much more difficult to hide. Fraud is also limited. The system monitored the traceability of dairy products inside a Romanian farm. Data about fodder provider and quality, cow productive performances and health and dairy products process were obtained and analyzed by students who will become specialists at all the levels of the food chain. Blockchain is the technology which in case of a dairy products contamination, the origin of the farm is traced in just a couple of seconds. In this way just a batch of dairy products is removed from distribution, leading to the reduction of food waste.

## 1 INTRODUCTION

The traceability of products and their components is a major challenge in the food industry, but also for the raw material producers. Traceability is related to information management. When a product undergoes a processing process, the information relating to that product must be processed so as to maintain the link between the product and the information [1].

Blockchain [2] technology ensures the implementation of the traceability principle for products, linking raw materials, their origin, processing, distribution and location after marketing, facilitating the design and implementation of ISO requirements. Traceability is particularly important today, as food ingredients come from all over the world [3].

Dairy is ideal for blockchain technology because of the amount of data collected at each point of the supply chain, starting at the herd level. Dairy factories must be able to trace the products of origin through traceability systems, tracking, for example, the milk used in a dairy to the cow farm. Knowing these things, factories can instantly identify and solve problems in each phase, traceability systems based on the process of recording information [4].

At the farm, the data can be linked to a milk shipment. This data contains information about the origin of the milk and about the laboratory work in which it is used. Production practices can be linked to the data, taking into consideration the animal welfare activities, environmental practices, carbon footprint outcomes and antibiotic applications [5]. According to Ramachandran, this system can also help producers to find out about what happens to their milk once it leaves the farm and allow them to make adjustments according to what the final customer needs to enrich the value of their product [4].

In the next section are presented the details about the Blockchain technology and how it works. Section 3 describes the proposed system in which Blockchain is used for foodtraceability. The last section outlines the conclusions.

## 2 METHODOLOGY

For training the students, it is necessary to create an e-learning platform that provides farmers, processors in the dairy industry and consumers all information on dairy products based on the Blockchain technology. For the teachers, by adopting this new technology, smart learning activities can be verified, as well as they are traceable [6]. The educational institution management can imply the software solution to evaluate the teaching performance.

In the educational system, Blockchain can be involved in the case of an educational transfer based on credits, respectively the European Credit Transfer and Accumulation System (ECTS) [7]. In this way it is enhanced a globally ubiquitous virtual environment that breaks the language and the administrative obstacles.

Blockchain can be implemented using the Hyperledger Fabric [8] open source platform. Traditionally, a permissioned blockchain requires every peer from the network to execute every transaction which occurs, to maintain a ledger and to use consensus [9]. Due to these actions, scalability is not enhanced, as well as private transactions and contracts are not supported [10].

The Hyperledger Fabric platform provides modularity, scalability and security, being used for industrial Blockchain solutions like the London Stock Exchange Group, CLS financial institution, U. S. Food and Drug Administration, TenneT energy community, SAP [11]. The improvement provided by Hyperledger Fabric is that the peers are split into two groups, namely endorser and commiter. Along with the peers exist the consenters.

Inside a market consisting of sellers, buyers, shippers, banks and other participants, offers can be published by the sellers at a certain price for a specific buyer, while the others to buy at the standard price. The details of the proposed deals are confidential. If the participants are not part of the deal, then the transaction does not appear on their ledger [12]. The identity is found from the membership service and sends the transactions only to specific peers. Each peer triggers a result, the transaction being validated if both peers provide the same result.

The peers deliver the validated transactions back to the application which forwards it to the consensus for ordering, after which the transactions are sent to the peers and stored inside the ledgers [13]. More parties are involved inside a transaction, such as the authorities, shippers, banks, but they do not need to know the details about the deal's price. The supply chain needs the Blockchain pattern to manage confidential transactions without forwarding everything through a central authority [14].

The deals and the transactions are stored inside the NoSQL database, CouchDB [15], using the JSON format. In this way queries can be done fast on a big amount of data. The Hyperledger Fabric platform uses CouchDB as a database due to its key-value storage, as well as GoLevelDB [16], which is an embedded database.

## 3 RESULTS

Blockchain technology makes it possible to identify the farms from where the cow milk was used in production, as well as the medical treatments or food rations that were used inside the farm. Using the RFID tags, it can be tracked the entire path traveled by the raw material.

The dynamic QR code is then printed on the packaging. This QR code contains all the information issued by the RFID tags together with information that is gradually enriched by the different participants in the supply chain. This includes any treatment to which the milk was subjected and the specific characteristics of the product. These data, which cannot be modified, continue to be placed in blocks until the product reaches the store.

Consumers can scan the QR code on the packaging to get all the information about the product stored in the blockchain blocks. In order to ensure maximum transparency, customers can also check data about the used blockchain technology.

The proposed system uses Blockchain which implemented based on the Hyperledger Fabric [6] open source platform and the CouchDB database due to its REST API done over a secure HTTP.

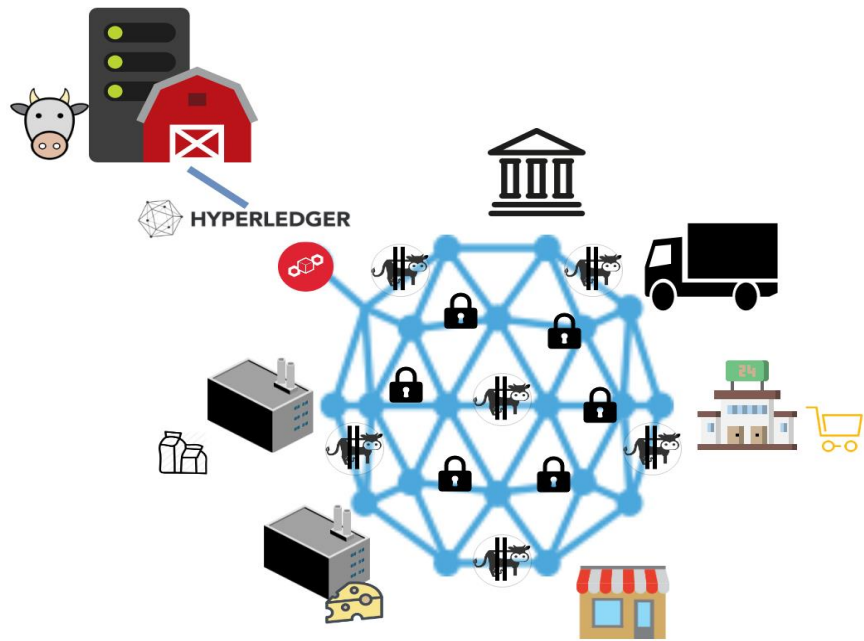

*Figure 1. Traceability based on Blockchain*

Fig. 1 depicts the traceability platform which is used to process, manage and control the transactions based on dairy products. The system consists of a distributed peer to peer network, where the peers of the Blockchain network are banks, transporters, dairy products processors, shops and supermarkets. The transactions are done online based on a virtual currency or by bank transfer. All the peers have a ledger for collecting tokens regarding the transactions.

Every time a dairy product undergoes a new process, the origin farm of the milk gets the token into the personal ledger. The transfer information is saved inside the blockchain, containing information about the source which has its true name, the receiver that is anonymized, the token which is the transaction product and the product identifier. Based on the blockchain address, the dairy farm is able to view the transactions without administrative obstacles. The process of assigning tokens is done through Hyperledger Fabric.

## 4   CONCLUSIONS

The advantages of Blockchain use are the speed of operation and automatic transmission of data to computers, precision due to the fact that the system provides a very low error rate, the system can provide complete, accurate and current data at any time, and eliminates the sources of errors which lead to additional costs to fix the errors. The use of this technique offers the possibility of using data automatically in industrial circuits. As a result, the storage of raw materials or finished products can be done more efficiently.

By applying this system will save time and staff. Over time, due to accurate inventory monitoring, their level of safety can be reduced, which is especially important for perishable products requiring special storage conditions.